\documentclass[usenatbib,twocolumn]{mn2e}
\usepackage{graphicx}
\usepackage{amsfonts}
\usepackage{epsfig,rotating,natbib,pstricks}
\usepackage{color}
\usepackage{amssymb,latexsym}
\usepackage{amsmath,wasysym}
\usepackage{siunitx}
\usepackage{verbatim}
\usepackage[T1]{fontenc}
\usepackage{aecompl}
\usepackage{float}
\usepackage{placeins}

\newcommand{\hi}{H{\sc i}}

\newcommand{\kms}{km s$^{-1}$}
\newcommand{\fescLA}{$f_\mathrm{esc}^\mathrm{Ly\alpha}$}
\newcommand{\fesc}{$f_\mathrm{esc}^\mathrm{LyC}$}
\newcommand{\vsep}{$\Delta v_{\mathrm{Ly\alpha}}$}

\usepackage{hyperref}
\hypersetup{
  colorlinks   = true,    
  urlcolor     = blue,    
  linkcolor    = blue,    
  citecolor    = blue      
}

\title[H{\sc i} Imaging of a Blueberry Galaxy]{H{\sc i} Imaging of a Blueberry Galaxy Suggests a Merger Origin}

\author[S. Dutta et al.]
{Saili Dutta$^{1}$, 
Apurba Bera$^{2,3}$, 
Omkar Bait$^{4}$, 
Chaitra A. Narayan$^{1}$, 
Biny Sebastian$^{5}$,
\newauthor{Sravani Vaddi$^{6}$}
\vspace{0.3cm}\\
$^{1}$ {National Centre for Radio Astrophysics, Tata Institute of Fundamental Research, Pune University Campus, Pune 411007, India} \\
$^{2}$ {International Centre for Radio Astronomy Research, Curtin University, Bentley, WA 6102, Australia}\\
$^{3}$ {Inter-University Centre for Astronomy and Astrophysics, Pune University Campus, Pune 411007, India} \\
$^{4}$ {Observatoire de Gen\`eve, Universit\'e de Gen\`eve, 51 Ch. des Maillettes, 1290 Versoix, Switzerland}\\
$^{5}$ {Department of Physics and Astronomy, University of Manitoba, Winnipeg, MB R3T 2N2, Canada} \\
$^{6}$ {Arecibo Observatory, NAIC, HC3 Box 53995, Arecibo, Puerto Rico, PR 00612, USA} 
}

\begin{document}

\maketitle

\begin{abstract}
Blueberry galaxies (BBs) are fainter, less massive, and lower redshift counterparts of the Green pea galaxies.
They are thought to be the nearest analogues of the high redshift Lyman Alpha (Ly$\alpha$) emitters.
We report the interferometric imaging of \hi\ 21 cm emission from a Blueberry galaxy, J1509+3731, at redshift, z = 0.03259, using the Giant Metrewave Radio Telescope (GMRT). 
We find that this Blueberry galaxy has an \hi\ mass of $M_{\text{H{\sc i}}} \approx 3\times 10^8 \, M_{\odot}$ and an \hi-to-stellar mass ratio $M_{\text{H{\sc i}}}/M_* \approx$ 2.4. 
Using SFR estimates from the H$\beta$ emission line, we find that it has a short \hi\ depletion time scale of $\approx 0.2$ Gyr, 
which indicates a significantly higher star-formation efficiency compared to typical star-forming galaxies at the present epoch. 
Interestingly, we find an offset of $\approx 2$ kpc between the peak of the H{\sc i} 21 cm emission and the optical centre which suggests a merger event in the past. Our study highlights the important role of mergers in triggering the starburst in BBs and their role in the possible leakage of Lyman-$\alpha$ and Lyman-continuum photons which is consistent with the previous studies on BB galaxies.

\end{abstract}



\section{Introduction}
\label{sec_intro}
In the current understanding of the $\Lambda$-Cold Dark Matter ($\Lambda$CDM) cosmology,
the structure formation in the Universe follows a hierarchical process.
This bottom-up structure formation starts with small over-densities, the dwarf galaxies.
Later massive galaxies grow from assembling these smaller galaxies
through mergers and other accretion activities.
A fraction of these low-mass young star-forming galaxies are the Lyman Alpha (Ly$\alpha$) emitters (LAEs); Lyman continuum (LyC) photons leaked by these LAEs
are thought to be the main contributors in reionizing the Universe
\citep{2015ApJ...803...34B, 2015ApJ...806...19D, 2017A&A...608A...6D, 2020MNRAS.495.3065D}.
Most known LAEs are high redshift, chemically less evolved, compact galaxies with 
low stellar mass, less dust, and a high ionization ratio. They are thought to be the progenitors of present-day Milky Way like galaxies, and provide an important probe to the earliest era of galaxy formation and the epoch of reionization \citep{2007ApJ...660.1023F, 2007ApJ...671..278G, 2012ApJ...745...12N, 2012ApJ...750L..36M, 2014MNRAS.442..900N, 2017ApJ...844..171Y, 2020ApJ...889..161N}. Detailed study of the inter-stellar medium (ISM) of LAEs is difficult because most LAEs are distant high redshift galaxies. However, identification of low redshift LAE-analogues allows us to indirectly probe the ISM properties of distant LAEs. 


The Galaxy Zoo project \citep{2008MNRAS.389.1179L} identifies a class of green-colored, 
small-sized objects from the Sloan Digital Sky Survey Data Release 7 \citep[SDSS DR7;][]{2009ApJS..182..543A},
known as the "Green Pea" galaxies (GPs). 
These highly star-forming (SFR $\sim 10 \; M_{\odot} yr^{-1}$) galaxies 
have high specific star formation rates (sSFR $\sim 10$ Gyr$^{-1}$)
for their stellar masses ($\log [M_*/M_{\odot}] \sim 8 - 10$),   
low metallicities ($12 + \log [O/H] \sim 8.05$)
and high 
[O${\; \textrm{III}}$]$ \lambda$5007/[O${\; \textrm{II}}$]$ \lambda$3727 (O32) ratios
\citep{2009MNRAS.399.1191C, 2010ApJ...715L.128A, 2012ApJ...749..185A, 2014ApJ...791L..19J, 2011ApJ...728..161I}. 
The bright green color of these compact galaxies arises from 
very strong [O$\; {\textrm{III}}$]$\lambda$5007 emission line
with large equivalent widths ($\sim 1000  $ \AA)
which falls into the SDSS `r' band.
After the discovery of these GPs, 
several studies 
have brought forward interesting similarities between
these GPs and the high redshift LAEs
\citep{2015ApJ...809...19H, 2016ApJ...820..130Y, 2017ApJ...844..171Y, 2020MNRAS.491..468I}.
\citet{2014ApJ...791L..19J} suggest that GPs
are optically thin systems, which makes it possible
for the LyC photons to escape.
The results from \citet{2015ApJ...809...19H} support this scenario indicating to 
low atomic hydrogen (H{\sc i}) column densities in GPs.
Outcomes of several studies on Ly$\alpha$ profiles, Ly$\alpha$ and LyC escape fractions of GPs 
\citep{2016ApJ...820..130Y, 2017ApJ...844..171Y, 2016Natur.529..178I, 2016MNRAS.461.3683I, 2018MNRAS.474.4514I, 2017MNRAS.471.2311L, 2019ApJ...874...52M}
strengthened the hypothesis that GPs are local analogues of high redshift LAEs. Importantly, it has been found that GPs can leak a significant fraction of LyC photons, making them similar to early star-forming galaxies believed to be responsible for reionization of the Universe \citep[e.g.,][]{2016Natur.529..178I, 2016MNRAS.461.3683I}. Several high-redshift ($z > 6$) galaxies, recently detected by James Webb Space Telescope (JWST), show properties very similar to GPs, thus increasing our confidence of them being good local analogues to early star-forming galaxies \citep{2022A&A...665L...4S, 2023ApJ...942L..14R, 2023arXiv230204298C, 2022arXiv221108255M}.

\citet{2017ApJ...847...38Y} (hereafter, Y17) present 40 starburst dwarf galaxies
from SDSS Data Release 12 \citep[SDSS DR12;][]{2015ApJS..219...12A}
catalogue\footnote{\emph{https://www.sdss4.org/dr12/}} with redshifts $z \leq 0.05$ 
and similar properties as GPs.
The [O$\; {\textrm{III}}$]$\lambda$5007 emission lines from these galaxies fall into the SDSS `g' band (because of their lower redshifts),
giving blue colours to the composite images, and leading to the name ``Blueberry'' Galaxies (BBs).
Later it has become a convention to refer all these low redshift analogues of LAEs as "Green Pea" galaxies.
However, we follow the traditional nomenclature and refer the lower redshift ($z < 0.1$) galaxies with blue-coloured composite images as "Blueberry" galaxies.
These BBs represent fainter, less massive, and 
lower redshift counterparts of the GPs, 
and are thought to be the nearest analogues of the high redshift LAEs.
A radio continuum study of some of these BBs shows 
suppression in the radio-based SFR compared to the optical SFR estimates \citep{2019ApJ...882L..19S}.


\citet{2019ApJ...872..145J} present a collection of 835 GPs from SDSS DR13 \citep{2017ApJS..233...25A} within the redshift range $0.011 < z < 0.411$.
\citet{2021ApJ...913L..15K} (hereafter, K21) select 40 galaxies from the sample of \citet{2019ApJ...872..145J} at $z<0.1$ and 
report the first detection of H{\sc i} 21 cm line emission in 19 GPs (we refer to these as BBs in this article as per the traditional nomenclature) along with estimates of their H{\sc i} masses using the Arecibo Telescope or the Green Bank Telescope (GBT).
They also find a lower H{\sc i} detection rate for 
galaxies with higher O32 ratios.  
\citet{2022ApJ...933L..11P} present the first interferometric H{\sc i} 21 cm imaging of a BB galaxy using the Giant Metrewave Radio Telescope (GMRT), which suggests a galaxy-galaxy merger event to be the reason for the starburst activity.

Here we present, H{\sc i} 21 cm imaging of a 
Blueberry galaxy, J1509+3731, at redshift\footnote{We assume a flat $\Lambda$CDM cosmology with ($\Omega_m, \Omega_{\Lambda}, h$) = (0.315, 0.685, 0.674) \citep{2020A&A...641A...6P},
where $h$ is the dimensionless Hubble parameter related to the
Hubble Constant $H_0 = 100 \, h \,$ km s$^{-1}$ Mpc$^{-1}$.} z = 0.03259. We describe the observations and data reduction process in Section \ref{sec_obs}. The observed \hi\ properties of the Blueberry galaxy are described in Section \ref{sec_res}. The implications of the observed \hi\ properties, in the context of high redshift LAEs, are discussed in Section \ref{sec_dis}. Finally, we summarize our results and conclusions in Section \ref{sec_sum}. 


\section{Observations and data analysis}
\begin{figure}
\centering
    \includegraphics[scale=0.65,trim={0cm 0cm 2.5cm 1.1cm},clip]{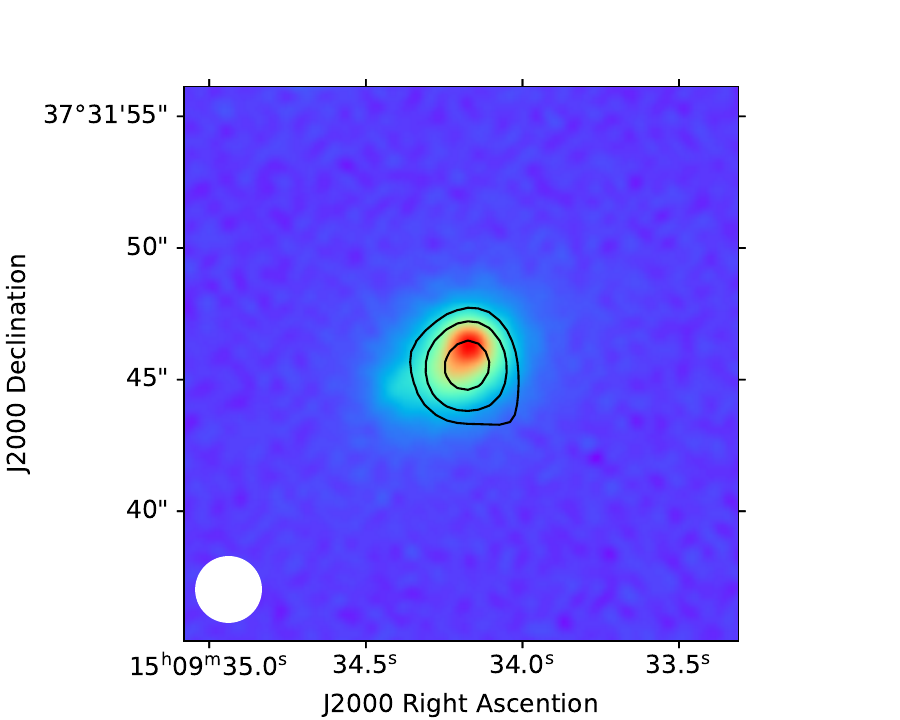}
  \caption{\textbf{Radio continuum emission from BB10:} The 1.4 GHz radio continuum contours are overlaid on 
  the DESI Legacy Survey g-band image. The contour levels are at 4, 8, and 16 times the RMS noise of the radio image. The synthesized beam of the radio image (2.5$\arcsec$ x 2.5$\arcsec$) is shown at the bottom left corner. The radio emission is resolved with a source size of $\approx 5 \arcsec$, about the same as the optical diameter of the galaxy.}
  \label{fig_cont}
\end{figure}
\label{sec_obs}
\subsection{The target galaxy}
We select J1509+3731 (hereafter BB10) with J2000 coordinates RA = 15$^h$09$^m$34.17$^s$, DEC = +37$^d$31$\arcmin$46.1$\arcsec$ from Y17 as the target galaxy for \hi\ 21 cm observation with the GMRT.
The redshift of BB10 is $z = 0.03259 \, (\pm 0.00001)$, which corresponds to a distance of 144 Mpc. 
The optical properties of BB10 are compiled in Table \ref{tab_prop}.
It has a high specific star formation rate ($\approx 13$ Gyr$^{-1}$) and a very high O32 ratio $(\approx 15.7)$.
In addition to these, its low metallicity and small optical size make it an ideal candidate to study as a local analogue of high redshift LAEs. 
Importantly, this galaxy shows strong Ly$\alpha$ emission and also evidence of Ly$\alpha$ absorption in the Hubble Space Telescope (HST) Cosmic Origins Spectrograph (COS) spectrum presented by \citet{2019ApJ...885...96J}. The Ly$\alpha$ emission features show two peaks with a large Ly$\alpha$ peak velocity separation of \vsep\ $= 400 \pm 27$~\kms and a weak Ly$\alpha$ escape fraction of \fescLA\ $=0.05 \pm 0.03$ 
\citep{2019ApJ...885...96J}.
Previously, K21 targeted this galaxy using GBT 
and provided an upper limit (3$\sigma$) on the H{\sc i} mass, 
$  M_{\textsc \hi} < 4.4 \times 10^8 \, M_{\odot}$.

\begin{table}
    \centering
    \begin{tabular}{cccc}
    \hline
    \hline
     $M_{*}$  & SFR  & O32 & $12 + \log [O/H]$ \\
      $\times \; 10^8 \; M_{\odot}$ & $M_{\odot} \;$ yr$^{-1}$ & &  \\
     \hline
    1.26 & 1.61 & 15.67 &  7.87\\
    \hline
    \end{tabular}
    \caption{The properties of J1509+3731 (BB10) from \citet{2017ApJ...847...38Y}. 
    The columns are i)stellar mass, ii) star formation rate, iii) O32 ratio, and iv) metallicity.}
    \label{tab_prop}
\end{table}

\begin{figure*}
\centering
    \includegraphics[scale=0.62,trim={0cm -0.1cm 2.5cm 1.2cm},clip]{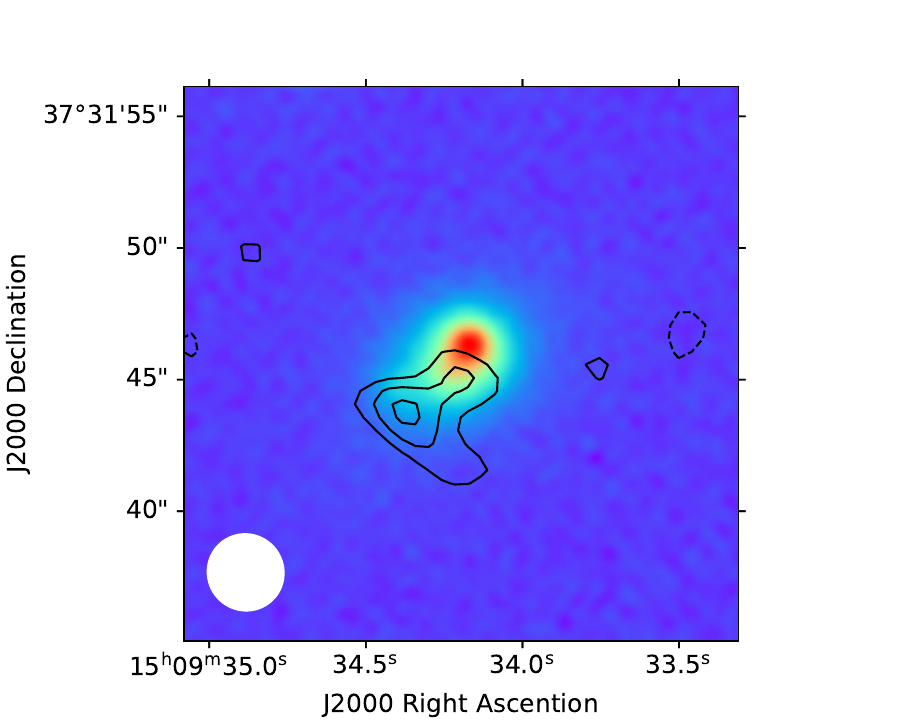}
    \includegraphics[scale=0.62,trim={0.4cm 0.35cm 0cm 0cm},clip]{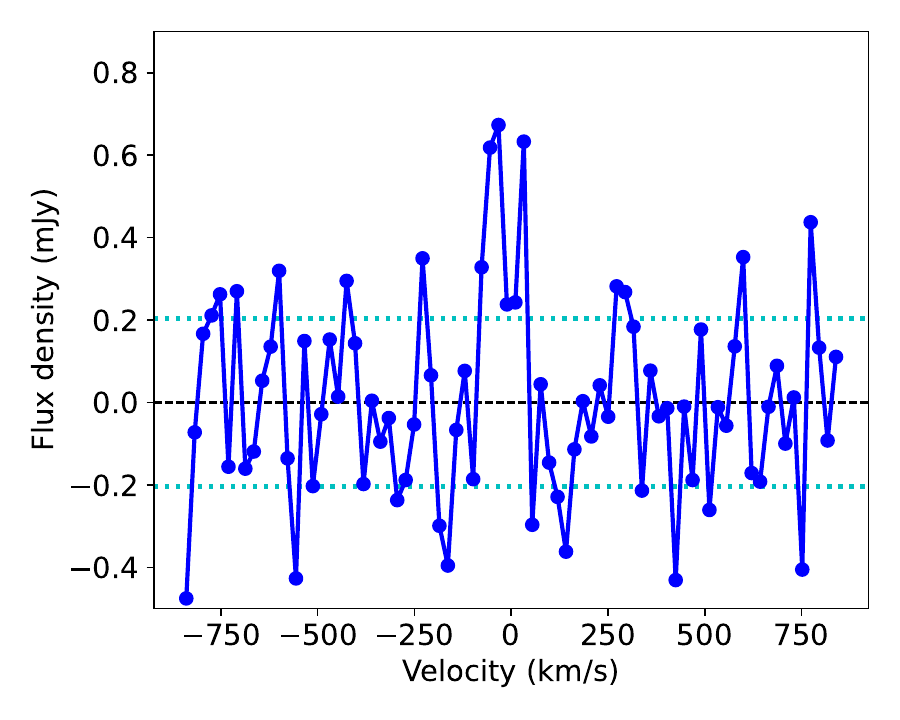}
  \caption{\textbf{[A, left] \hi\ 21 cm emission map of BB10 at a resolution of $\mathbf{2.9\arcsec}$:} The \hi\ contours, corresponding to the emission map obtained by averaging six central velocity channels within the FWHM of the \hi\ 21 cm signal, are overlaid on the DESI Legacy Survey g-band image.  The contour levels are at 2, 3, and 4 times the RMS noise in the \hi\ 21 cm emission map. The negative contours are shown in dashed lines. The synthesized beam is shown at the bottom left corner.
  \textbf{[B, right] Spatially integrated \hi\ 21 cm emission spectrum of BB10 at a velocity resolution of 21.8 km s$^{-1}$:} The \hi\ 21 cm spectrum has been integrated over the region within the outermost contour on the left plot. The dotted line shows the spectral RMS noise.}
  \label{fig_spec}
\end{figure*}

\subsection{Observations and Data Analysis}
The target galaxy, BB10, was observed with the band-5 (1000-1450 MHz) system of the upgraded GMRT (uGMRT), under the observing proposal DDTC095 (PI: Chaitra A. Narayan), for a total on-source time of $\sim$9 hrs. The observations were carried out with a bandwidth of 100 MHz (divided into 8192 channels) centered at 1380.00 MHz, using the GMRT Wideband Backend (GWB). Radio sources 3C286 and 1602+334 were used as the flux and the phase calibrator, respectively.
The central square baselines, for the GWB system, were affected by a power-offset\footnote{A detailed description of the issue can be found at \url{http://gmrt.ncra.tifr.res.in/gmrt\_users/help/csq\_baselines.html}.}.
Hence the central square baselines were excluded from our analysis.

\subsection{Radio continuum imaging}


Radio interferometric imaging was carried out using the CASA (Common Astronomy Software Applications) software package
\citep{2007ASPC..376..127M}. 
After gain calibration and bandpass calibration using the flux calibrator, 
visibility data were inspected for the presence of radio frequency interference (RFI). 
Following the excision of RFI-affected data, visibilities were averaged to 781 kHz frequency channels for continuum imaging.
Channels within $\pm 200$ km s$^{-1}$ of the expected location of the \hi\ 21 cm line were not used for continuum imaging. 
The continuum image was made using the task {\sc tclean}, and was used to self-calibrate the visibility data. 
The final continuum image has an RMS noise of 12.4 $\mu$Jy beam$^{-1}$ and an angular resolution of 2.5$\arcsec$ x 2.5$\arcsec$.

Radio continuum emission from BB10 was detected in our continuum image, as shown in Figure~\ref{fig_cont} where we plot the 1.4 GHz radio continuum contours on the optical g-band image \footnote{\url{https://www.legacysurvey.org/}} from Dark Energy Spectroscopic Instrument (DESI) Legacy imaging survey \citep{2019AJ....157..168D}. 
We used the \textsc{CASA} task \textsc{IMFIT} to estimate the properties of the emission source by fitting a 2-dimensional elliptical Gaussian to the spatial brightness distribution. A detailed description of the task, and the algorithms used, can be found in the \textsc{CASA} documentation \footnote{\url{https://casadocs.readthedocs.io/en/latest/api/tt/casatasks.analysis.imfit.html}}. The uncertainties on the flux density and the sky position of the peak are estimated using the prescription of \citet{condon97}.

The radio continuum emission is marginally resolved\footnote{The peak flux density and the integrated flux density of the source differ by $\approx 2.5 \sigma$. It has a deconvolved size of 1.6\arcsec $\times$ 1.0\arcsec, which is smaller than the synthesized beam. We verified these estimates by performing source extraction using a different software package \citep[PyBDSF;][]{2015ascl.soft02007M}, which yielded consistent results.} with an integrated 1.4 GHz flux density of 340 $\pm$ 29 $\mu$Jy. Our estimate of the (integrated) flux density is consistent with its previously published measurement \citep[$345 \pm 16 \; \mu$Jy;][]{2019ApJ...882L..19S}. The peak flux density of the radio continuum emission (at an angular resolution of 2.5\arcsec) is 261 $\mu$Jy beam$^{-1}$, detected at a statistical significance of (S/N)$_{\rm peak}$ = 21 (i.e. peak / RMS noise).

The uncertainties on the sky position of the peak of the emission source is given by \citep{condon97,1998AJ....115.1693C} 

\begin{equation}
    \rm \Delta RA \approx \Delta DEC \approx 
    \frac{\Theta_{\rm FWHM}}{2 (S/N)_{peak}}
    \label{eqn_dpeak}
\end{equation}

where $\Theta_{\rm FWHM}$ is the FWHM of the synthesized beam and $\rm (S/N)_{peak}$ is the ratio of the peak flux density (not the spatially integrated flux density) to RMS noise. For example, the sky position of a source with a peak S/N = 5 can be measured with an accuracy of approximately a tenth of the FWHM of the synthesized beam in each of RA and DEC \citep{1998AJ....115.1693C}.

The peak of the radio continuum emission associated with BB10 is offset by $0.7\arcsec \pm 0.1 \arcsec$ (0.49 $\pm$ 0.06 kpc) from the peak of optical emission (see Figure~\ref{fig_cont}). The uncertainties on the position of the radio peak is consistent with the relation in Equation~\ref{eqn_dpeak}. To investigate any systematic spatial offset in our radio image, we compared the positions of bright compact sources in our continuum image (within the FWHM of the GMRT primary beam) with their reference positions in the NRAO VLA Sky Survey \citep[NVSS\footnote{\url{https://www.cv.nrao.edu/nvss/}};][]{1998AJ....115.1693C} catalogue. The individual offsets along RA and DEC, i.e. $\delta {\rm RA = RA - RA_{NVSS}}$ and $\delta {\rm DEC = DEC - DEC_{NVSS}}$, are randomly distributed with both positive and negative values, around their median values of $\delta {\rm RA_{median}} = 0.1\arcsec$ and $\delta {\rm DEC_{median}} = 0.04 \arcsec$. Note that $\delta {\rm RA_{median}}$ and $\delta {\rm DEC_{median}}$, the median values of the offsets along RA and DEC, represent the systematic offset between our radio image and NVSS. This systematic offset is significantly smaller than the offset between the radio and the optical peaks, and hence we do not apply any correction for the systematic offset.

\subsection{Spectroscopic imaging}


The gain solutions from self-calibration were applied to visibilities at the native frequency resolution (before channel averaging). 
After another round of RFI excision, the calibrated visibilities were used to make \hi\ 21 cm data cube with a velocity resolution of 21.8 \kms. 
The \hi\ data cube was made using visibilities with a UV upper limit of 60 k$\lambda$ (and Briggs weighting scheme, with robust = 1.0), which yielded a spatial resolution of $2.9 \arcsec \times 2.9 \arcsec$. The RMS noise at this angular resolution, per 21.8 \kms\ velocity channel, is $\approx 170  \; \mu$Jy beam$^{-1}$.
A spectral baseline was subtracted from each spatial pixel in the cube, where the spectral baseline was obtained by fitting a second-order polynomial to the spectra excluding channels within $\pm 200$ km s$^{-1}$ of the expected location of the \hi\ 21 cm line. 
Additional data cubes were made at lower spatial resolutions, using smaller UV upper limits. For example, the cube made using visibilities with a UV upper limit of 15 k$\lambda$ (and Briggs weighting scheme, with robust = 1.0) has a spatial resolution of $7.4 \arcsec \times 6.5 \arcsec$ and an RMS noise of $\approx  280 \; \mu$Jy beam$^{-1}$ per 21.8 \kms\ velocity channel.
Both the data cubes are CLEANed down to 1.5 times the rms noise in each channel.

As mentioned earlier, the central square baselines (< 5 k$\lambda$) of GMRT were not used for making the spectral cubes. The largest angular scale probed by these cubes is thus $\approx 40 \arcsec$. Hence, the H{\sc i} 21 cm emission can be reliably mapped up to angular scales of $\lesssim 20 \arcsec$ (spatial scales of $\lesssim 15$ kpc at the redshift of BB10).

\begin{figure}
\centering
    \includegraphics[scale=0.62,trim={0cm -0.1cm 2.4cm 1.2cm},clip]{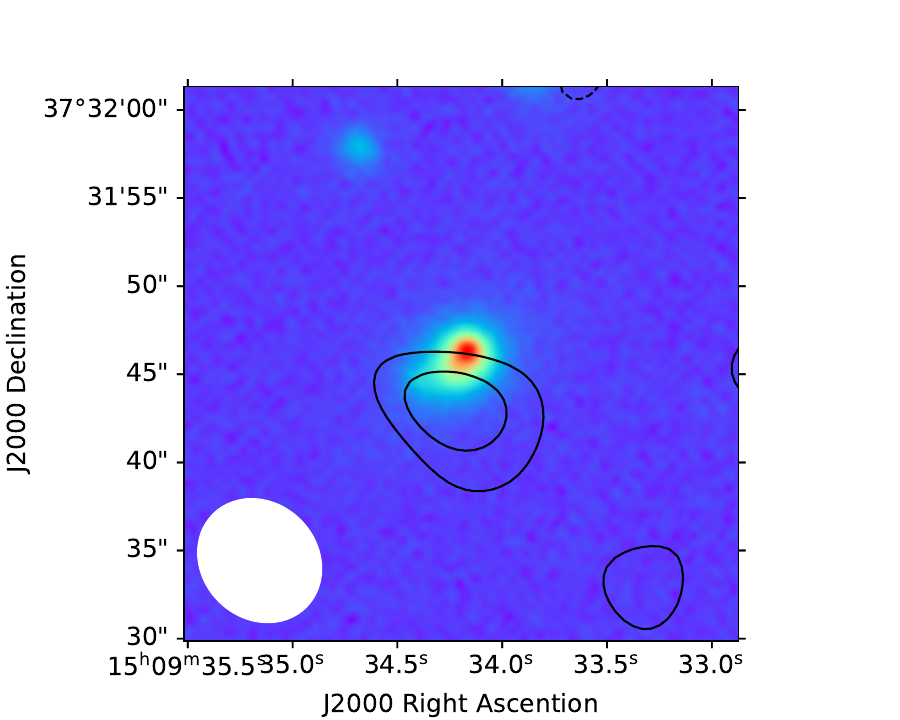}
\caption{The \hi\ contours at a resolution of $7.4\arcsec \times 6.5\arcsec$ overlaid on the DESI Legacy Survey g-band image: The contour levels are at 2, and 3 times the RMS noise in the \hi\ 21 cm map at this resolution. The negative contours are shown in dashed lines. The synthesized beam is shown at the bottom left corner.}
  \label{fig_maps}
\end{figure}

\section{H{\sc i} properties of BB10}
\label{sec_res}

\hi\ 21 cm emission from BB10 is detected in our data cube (at $2.9 \arcsec \times 2.9 \arcsec$ angular resolution). The \hi\ 21 cm spectrum was obtained by spatially integrating over the source. 
The \hi\ 21 cm emission (moment-0) map was obtained by averaging six velocity channels within the FWHM of the \hi\ 21 cm line (also above $1\sigma$).
Figure \ref{fig_spec}[A] shows the \hi\ 21 cm emission map of BB10, overlaid on the DESI Legacy Survey g-band image, and 
Figure \ref{fig_spec}[B] 
shows its \hi\ 21 cm spectrum at a velocity resolution of 21.8 km s$^{-1}$. 
The integrated H{\sc i} 21 cm flux density is $S_{21} = 60 \pm 11$ mJy km s$^{-1}$, 
which yields an H{\sc i} mass of $M_{\text{H{\sc i}}} = (3.12 \pm 0.57) \times 10^8 \, M_{\odot}$. 
At an angular resolution of $2.9 \arcsec \times 2.9 \arcsec $, the \hi\ 21 cm emission map has a 2$\sigma$ column density sensitivity of $N_{\text{H{\sc i}}} = 1.9 \times 10^{21}$ cm$^{-2}$.

The \hi\ 21 cm spectra and emission maps obtained from data cubes made at coarser spatial resolution yield \hi\ masses consistent with the above estimate. 
The \hi\ 21 cm emission map at a spatial resolution of $7.4 \arcsec \times 6.5 \arcsec$ is shown in Figure \ref{fig_maps}
, which has a 2$\sigma$ column density sensitivity of $N_{\text{H{\sc i}}} = 0.7 \times 10^{21}$ cm$^{-2}$.
Table \ref{tab_res} compiles the details of integrated H{\sc i} 21 cm flux density, H{\sc i} mass, and 2$\sigma$ H{\sc i} column density for different beam sizes.
As the central square baselines are not included, 
it could not be possible to consider a bigger synthesized beam and make data cubes with coarser resolution. Therefore, our estimate for the H{\sc i} mass is formally a lower limit of the actual H{\sc i} mass of the galaxy.

Due to the lack of short baselines, our observations are not sensitive to H{\sc i} emission on spatial scales $\gtrsim$ 15 kpc. However, we note that the non-detection of H{\sc i} 21cm emission from BB10 with the GBT by K21 and the associated upper limit on its H{\sc i} mass suggest that the galaxy does not have a significant amount of H{\sc i} at larger spatial scales. The GBT upper limit  --- $M_{\text{H{\sc i}}} < 4.4 \times 10^8 M_{\odot}$ --- and the (remarkably) tight H{\sc i} mass–size relation \citep{2016MNRAS.460.2143W} suggests a H{\sc i} disk size (diameter) of $d_{HI} \lesssim 15$ kpc. It thus appears unlikely that the actual H{\sc i} mass of BB10 is significantly larger than our estimate.

The \hi\ depletion timescale, $t_{\text{dep}} = M_{\text{H{\sc i}}}/SFR$, indicates how long a galaxy can continue forming stars without refilling its gas reservoir. BB10 has a short \hi\ depletion timescale of $\approx 0.2$ Gyr.

\begin{table}
    \centering
    \begin{tabular}{cccc}
    \hline
    \hline
     Beam  & $S_{\text{int}}$  & $M_{\text{H{\sc i}}}$ & $N_{\text{H{\sc i}}}$\\
     ($\arcsec \times \arcsec$) & (mJy km s$^{-1}$) & ($\times \; 10^8 \; M_{\odot}$) & ($\times 10^{21}$ cm$^{-2}$) \\
     \hline
    2.9 $\times $ 2.9 & 60 $\pm$ 11 &  3.12 $\pm$ 0.57 &  1.9\\
    7.4 $\times $ 6.5 & 68 $\pm$ 14 &  3.53 $\pm$ 0.73 & 0.7 \\
     \hline
    \end{tabular}
    \caption{The values of H{\sc i} mass ($M_{\text{H{\sc i}}}$) and 
    2$\sigma$ H{\sc i} column density sensitivity (integrated over the \hi\ 21 cm emission line with velocity width $\approx$ 130.8 \kms)
    at different angular resolutions.}
    \label{tab_res}
\end{table}

An offset of $2.88\arcsec \pm 0.70\arcsec$, i.e. $1.94 \pm 0.48$ kpc, 
is seen between the peak of the optical emission and 
the peak of H{\sc i} distribution\footnote{The peak of the H{\sc i} distribution was localized by fitting a 2D Gaussian to the emission source in the moment-0 map.}
The H{\sc i} contours are elongated along the optical extension(see Figure \ref{fig_spec}[A]).
The offset is seen in the \hi\ 21 cm maps at coarser spatial resolutions as well. 
This offset is significantly larger than the offset  between the peak of optical emission and that of radio continuum emission (see Figure \ref{fig_cont}). 


\begin{figure*}
\centering
    \includegraphics[scale=0.62,trim={0cm 0cm 0.0cm 0.0cm},clip]{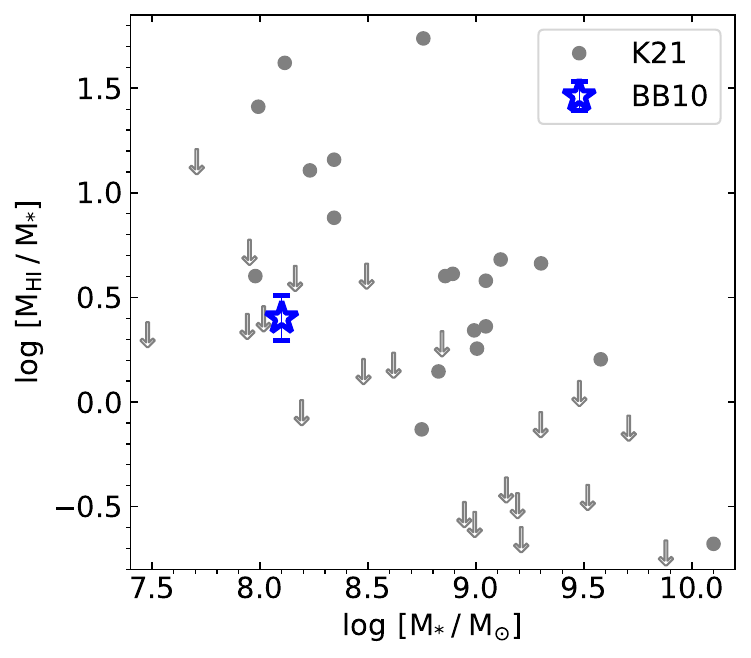}
    \includegraphics[scale=0.62,trim={0cm 0cm 0.0cm 0.0cm},clip]{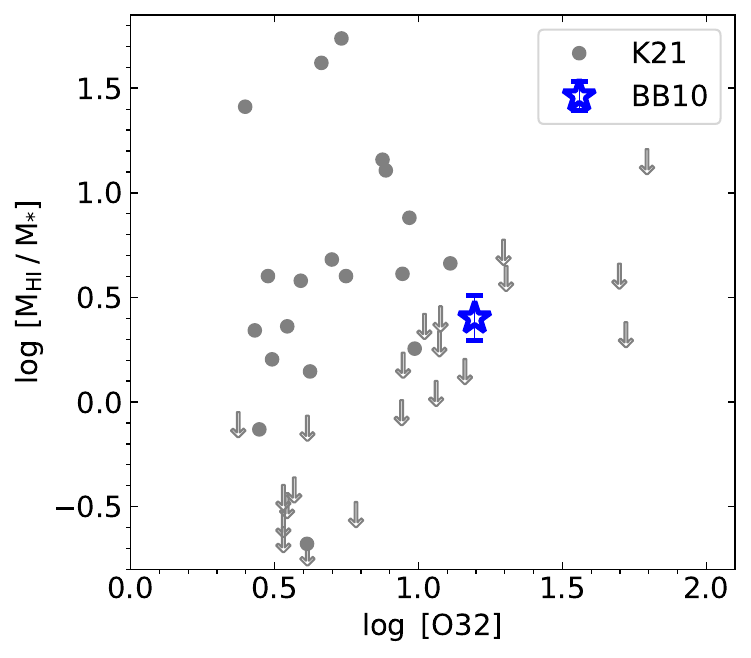}
    \includegraphics[scale=0.62,trim={0cm 0cm 0.0cm 0.0cm},clip]{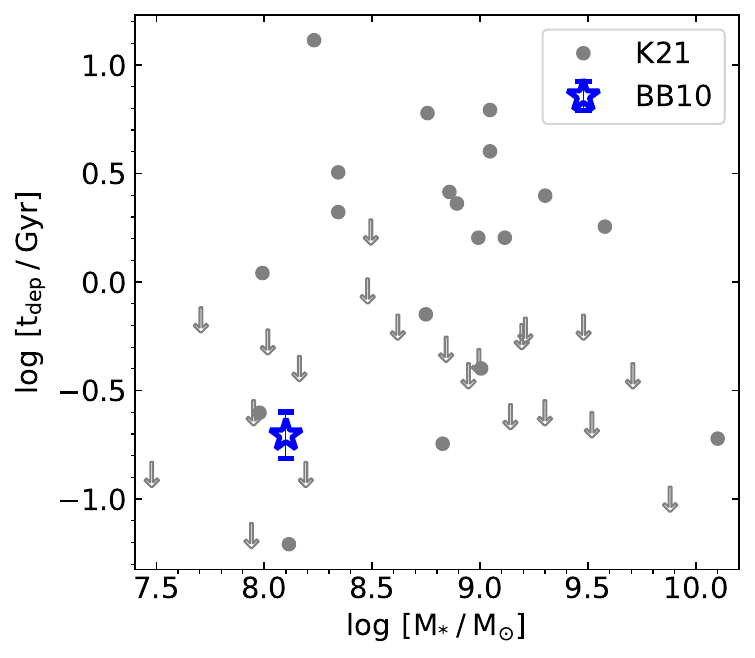}
    \includegraphics[scale=0.62,trim={0cm 0cm 0.0cm 0.0cm},clip]{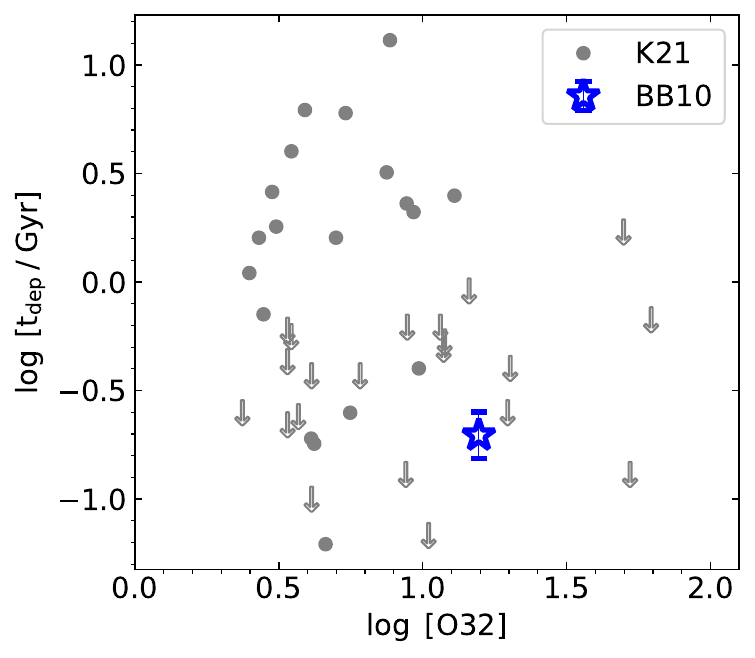}
  \caption{\textbf{$\mathbf{M_{\text{H{\sc i}}} / M_*}$ [top panels] and \hi\ depletion time scale [bottom panels] of galaxies from K21 and BB10 as functions of their stellar mass [left] and O32 
  [right]:} The grey filled circles show \hi\ 21 cm detections of Blueberry galaxies and the grey arrows show 3$\sigma$ upper limits for non-detections from \citet{2021ApJ...913L..15K}. The blue star represents BB10.}
  \label{fig_gas}
\end{figure*}

\section{Discussion}
\label{sec_dis}

We compare the H{\sc i} properties of BB10 with the BBs from K21. We note that BB10 was also included in the sample of K21.
Out of a sample of 40 BBs, K21 provide estimates of \hi\ masses for 19 H{\sc i} detections and upper limits for the remaining 21 including BB10.
Our H{\sc i} mass estimate for BB10 is consistent with the upper limit provided by K21 ($4.4 \times 10^8 \, M_{\odot}$).
Figure \ref{fig_gas} shows the H{\sc i}-to-stellar mass ratio, $M_{\text{H{\sc i}}}/M_*$, 
and the \hi\ depletion timescale, $t_{\text{dep}} = M_{\text{H{\sc i}}}/SFR$, 
against the stellar mass and O32 ratio for the BBs from the sample of K21 along with BB10.
The grey filled circles represent the H{\sc i} detections and the grey downward arrows show the upper limits for the non-detections from K21. In all four panels, BB10 is shown as a blue star. 
As can be seen in the top left and bottom left panels of Figure \ref{fig_gas}, BB10 has a somewhat similar H{\sc i}-to-stellar mass ratio and \hi\ depletion timescale as the BBs from K21 with similar stellar masses. However, it is on the lower side of the  H{\sc i}-to-stellar mass ratio and \hi\ depletion timescale distribution. 
The top right and bottom right panels of Figure \ref{fig_gas} show, as also noted by K21, that galaxies with higher O32 values have preferentially lower H{\sc i}-to-stellar mass ratio, and thus also typically lower depletion timescales. Our estimates of the H{\sc i}-to-stellar mass ratio and \hi\ depletion timescale of BB10 are consistent with this scenario. We note that BB10 has a higher O32 ratio than the other BBs with \hi\ 21 cm detections in the K21 sample. However, more measurements of the \hi\ properties of galaxies in the low $M_*$ - high O32 space are needed to make a robust conclusion.

\citet{2022ApJ...933L..11P} find H{\sc i} 21 cm emission from a broken ring-like structure around a BB galaxy
and suggest a merger event, with a companion galaxy separated by $\approx 4.7 $ kpc,
to be responsible for the starburst activity in that BB galaxy. Recently, \citet{2023MNRAS.tmp.3772L} present H{\sc i} imaging of a well-studied local LyC emitter (LCE), Haro 11. Both these studies highlight that mergers can cause a significant displacement of the neutral gas which can facilitate the escape of Lyman-$\alpha$ and possibly also Lyman-continuum photons. 

We do not find any clear evidence of a close companion galaxy (or a merging galaxy) from the DESI Legacy Survey \citep{2019AJ....157..168D} optical image of BB10.
However, as mentioned earlier, we do notice a clear spatial offset between the peak of the \hi\ 21 cm emission and the star-forming region (the brightest part of the optical emission). In addition, the H{\sc i} distribution extends over the star-forming region as well.
We note that optical emission in the DESI image also appears asymmetric as can be seen in Figure~\ref{fig_cont}.
While the brightest part of the optical emission is offset from the densest part of the H{\sc i} spatial distribution,
an extended `tail' of the optical emission coincides with the peak of the H{\sc i} distribution (see Figure~\ref{fig_spec}[A]).
Both of these might be caused by a dwarf-dwarf merger. And in that case, BB10 is perhaps in a late-stage merger where the two galaxies have nearly coalesced with each other. Interestingly, the COS-NUV image of BB10, which has a higher spatial resolution than the DESI image, shows relatively extended emission with two or more UV emitting knots within the bright optical emission \citep[see Figure~3 in][]{2019ApJ...885...96J}.  It is well known that mergers can trigger a starburst \citep[e.g.,][]{2014ApJ...789L..16L, 2015ApJ...807L..16K, 2015ApJ...805....2S, 2021MNRAS.503.3113M}. Thus a dwarf-dwarf merger scenario can altogether explain the origin of the offset in \hi\ 21 cm emission, multiple UV knots, and the origin of the starburst in BB10.

Finally, compact starbursts can have strong outflows due to stellar/supernova-driven feedback \citep[e.g.,][]{2009ApJ...692..187W, 2011ApJ...730....5H, 2017A&A...605A..67C}. Such outflows can push significant amounts of gas from the star-forming knot of BB10 leading to a paucity of gas in that region.

Using a sample of $z \sim 0.3$ high O32 star-forming galaxies, \citet{2018MNRAS.474.4514I} propose an empirical relation between \vsep{} and the escape fraction of LyC photons (\fesc{}). Further, \citet{2022ApJ...930..126F} using a multi-parameter analysis finds that \vsep{} is the best in-direct indicator of \fesc{}. BB10 has properties which lie within the range of galaxies studied in \citet{2018MNRAS.474.4514I}, in terms of the O32 ratio, stellar mass, metallicity, SFR, etc. We can thus get an indirect estimate of \fesc\ from the observed \vsep\ $=400 \pm 27$ \kms from \citet{2019ApJ...885...96J} and the empirical relation from \citet{2018MNRAS.474.4514I}. We find an indirect \fesc\ $\approx$ 0.034 for BB10. Thus BB10 is a weak Ly$\alpha$ and possibly a weak LyC leaker.
Indeed mergers can lead to favourable conditions such as triggering an intense starburst (and thus increasing the intrinsic LyC production) and simultaneously displacing a significant amount of \hi{} gas through tidal interactions to open low column density channels which can facilitate the leakage of Ly$\alpha$ and possibly LyC photons \citep{2024NatAs...8..384W}.
However, H{\sc i} 21 cm imaging of a sample of BBs would be needed to help us obtain a better understanding of the role of galaxy mergers, in particular the complex role of orientation and geometrical effects, in the escape of LyC photons. Targeted simulations of merging galaxies with realistic feedback will also be helpful in understanding the role of mergers in LyC escape from these young star-forming galaxies \citep[e.g.,][]{2015MNRAS.446.2038R}.

\section{Conclusion \& Summary}
\label{sec_sum}

In this article, we present the results obtained from H{\sc i} 21 cm imaging of a Blueberry galaxy at a redshift of $z = 0.03259$, using uGMRT band-5 receivers.
This Blueberry galaxy has an H{\sc i} mass of $M_{\text{H{\sc i}}} = (3.12 \pm 0.57) \times 10^8 \, M_{\odot}$, and an \hi-to-stellar mass ratio $M_{\text{H{\sc i}}}/M_* \approx$ 2.4. Using SFR estimates from the H$\beta$ emission line \citep{2017ApJ...847...38Y}, 
we find that it has a short \hi\ depletion time scale of $\approx 0.2$ Gyr. 
Comparing the H{\sc i} 21 cm emission map with the DESI Legacy Survey optical image of BB10, we find an offset of 
$(1.94 \pm 0.48)$ kpc between the centres of H{\sc i} 21 cm emission and star-forming region. 
Such an offset could be a sign of a merger event which can also trigger a starburst. Our study along with \citet{2022ApJ...933L..11P} and \citet{2023MNRAS.tmp.3772L} highlight the role of dwarf galaxy mergers in the leakage of ionizing photons, and thus their role in cosmic reionization. 

\section*{ACKNOWLEDGMENTS}
We are grateful to the anonymous reviewer and the editor for feedback and suggestions on the manuscript which have improved the article significantly.
We thank the staff of the GMRT who have made these observations possible. 
The GMRT is run by the National Center for Radio Astrophysics of the 
Tata Institute of Fundamental Research. O. B. is supported by the {\em AstroSignals} Sinergia Project funded by the Swiss National Science Foundation. S.D and C.A.N. also acknowledge the Department of Atomic Energy for funding support, under project 12-R\&D-TFR-5.02-0700.

The Legacy Surveys consist of three individual and complementary projects: the Dark Energy Camera Legacy Survey (DECaLS; Proposal ID \#2014B-0404; PIs: David Schlegel and Arjun Dey), the Beijing-Arizona Sky Survey (BASS; NOAO Prop. ID \#2015A-0801; PIs: Zhou Xu and Xiaohui Fan), and the Mayall z-band Legacy Survey (MzLS; Prop. ID \#2016A-0453; PI: Arjun Dey). DECaLS, BASS and MzLS together include data obtained, respectively, at the Blanco telescope, Cerro Tololo Inter-American Observatory, NSF’s NOIRLab; the Bok telescope, Steward Observatory, University of Arizona; and the Mayall telescope, Kitt Peak National Observatory, NOIRLab. Pipeline processing and analyses of the data were supported by NOIRLab and the Lawrence Berkeley National Laboratory (LBNL). The Legacy Surveys project is honored to be permitted to conduct astronomical research on Iolkam Du’ag (Kitt Peak), a mountain with particular significance to the Tohono O’odham Nation.

NOIRLab is operated by the Association of Universities for Research in Astronomy (AURA) under a cooperative agreement with the National Science Foundation. LBNL is managed by the Regents of the University of California under contract to the U.S. Department of Energy.

This project used data obtained with the Dark Energy Camera (DECam), which was constructed by the Dark Energy Survey (DES) collaboration. Funding for the DES Projects has been provided by the U.S. Department of Energy, the U.S. National Science Foundation, the Ministry of Science and Education of Spain, the Science and Technology Facilities Council of the United Kingdom, the Higher Education Funding Council for England, the National Center for Supercomputing Applications at the University of Illinois at Urbana-Champaign, the Kavli Institute of Cosmological Physics at the University of Chicago, Center for Cosmology and Astro-Particle Physics at the Ohio State University, the Mitchell Institute for Fundamental Physics and Astronomy at Texas A\&M University, Financiadora de Estudos e Projetos, Fundacao Carlos Chagas Filho de Amparo, Financiadora de Estudos e Projetos, Fundacao Carlos Chagas Filho de Amparo a Pesquisa do Estado do Rio de Janeiro, Conselho Nacional de Desenvolvimento Cientifico e Tecnologico and the Ministerio da Ciencia, Tecnologia e Inovacao, the Deutsche Forschungsgemeinschaft and the Collaborating Institutions in the Dark Energy Survey. The Collaborating Institutions are Argonne National Laboratory, the University of California at Santa Cruz, the University of Cambridge, Centro de Investigaciones Energeticas, Medioambientales y Tecnologicas-Madrid, the University of Chicago, University College London, the DES-Brazil Consortium, the University of Edinburgh, the Eidgenossische Technische Hochschule (ETH) Zurich, Fermi National Accelerator Laboratory, the University of Illinois at Urbana-Champaign, the Institut de Ciencies de l’Espai (IEEC/CSIC), the Institut de Fisica d’Altes Energies, Lawrence Berkeley National Laboratory, the Ludwig Maximilians Universitat Munchen and the associated Excellence Cluster Universe, the University of Michigan, NSF’s NOIRLab, the University of Nottingham, the Ohio State University, the University of Pennsylvania, the University of Portsmouth, SLAC National Accelerator Laboratory, Stanford University, the University of Sussex, and Texas A\&M University.

BASS is a key project of the Telescope Access Program (TAP), which has been funded by the National Astronomical Observatories of China, the Chinese Academy of Sciences (the Strategic Priority Research Program “The Emergence of Cosmological Structures” Grant \# XDB09000000), and the Special Fund for Astronomy from the Ministry of Finance. The BASS is also supported by the External Cooperation Program of Chinese Academy of Sciences (Grant \# 114A11KYSB20160057), and Chinese National Natural Science Foundation (Grant \# 12120101003, \# 11433005).

The Legacy Survey team makes use of data products from the Near-Earth Object Wide-field Infrared Survey Explorer (NEOWISE), which is a project of the Jet Propulsion Laboratory/California Institute of Technology. NEOWISE is funded by the National Aeronautics and Space Administration.

The Legacy Surveys imaging of the DESI footprint is supported by the Director, Office of Science, Office of High Energy Physics of the U.S. Department of Energy under Contract No. DE-AC02-05CH1123, by the National Energy Research Scientific Computing Center, a DOE Office of Science User Facility under the same contract; and by the U.S. National Science Foundation, Division of Astronomical Sciences under Contract No. AST-0950945 to NOAO.

\section*{Data Availability}
Raw data used in this work are available in the GMRT Online Archive (\emph{http://naps.ncra.tifr.res.in/goa/}) under proposal ID: DDTC095. Reduced data are available with the corresponding author and will be shared on reasonable request.


\end{document}